# Learning from Galileo's errors

### Enrico Bernieri

Four hundred years after its publication, Galileo's masterpiece *Sidereus Nuncius* is still a mine of useful information for historians of science and astronomy. In his short book Galileo reports a large amount of data that, despite its age, has not yet been fully explored. In this paper Galileo's first observations of Jupiter's satellites are quantitatively re-analysed by using modern planetarium software. All the angular records reported in the *Sidereus Nuncius* are, for the first time, compared with satellites' elongations carefully reconstructed taking into account software accuracy and the indeterminacy of observation time. This comparison allows us to derive the experimental errors of Galileo's measurements and gives us direct insight into the effective angular resolution of Galileo's observations. Until now, historians of science have mainly obtained these indirectly and they are often not correctly estimated. Furthermore, a statistical analysis of Galileo's experimental errors shows an asymmetrical distribution with prevailing positive errors. This behaviour may help to better understand the method Galileo used to measure angular elongation, since the method described in the *Sidereus Nuncius* is clearly wrong.

In 1964, three hundred and fifty-four years after the publication of the *Sidereus Nuncius* (Venice 1610), the Belgian astronomer Jean Meeus, analysing Galileo's observations of Jupiter's satellites, was the first to show that all the four 'new' satellites were potentially visible at the time of the first observation on 1610 January 7.[1] Galileo could not resolve satellites I and II, later named Io and Europa by Simon Mayr, because of the limits of his telescope. Meeus also showed that Galileo was unable to see the satellites when they were within about three equatorial Jupiter radii from the edge of the planet, giving a first direct indication of the resolution of Galileo's instrument. Meeus calculated Jovian satellite elongations at the time of Galileo's early observations by using his new computed tables, which allowed him to obtain the configuration of Jupiter's main satellites between the years 1600 and 2200.[2]

The historian of science Stillman Drake, one of the leading modern experts on Galileo, followed Meeus' work and, in an appendix to his 1983 translation and commentary on *Sidereus Nuncius*, analysed all 65 Jupiter observations, comparing the descriptions of the satellites' positions with their positions calculated by modern ephemerides.[3] Drake concluded that Galileo was generally able to resolve two satellites when they were separated by more than about 10 arcseconds. Unfortunately, it is difficult to verify Drake's calculations because he doesn't record either the ephemeris used or the exact times assumed for the observations.

The resolution and other parameters of Galileo's telescope have been and still are debated among historians of science. Since the telescope no longer exists, many historians based their hypotheses mainly on the analysis of other subsequent telescopes built by Galileo and not on the direct analysis of Galileo's data.[4,5,6] Moreover, until a few years ago, only professional astronomers were in a position to determine what Galileo had actually seen as compared with his recorded observations and his calculations, and it is not a task likely to interest them. On the other hand, probably most historians have incorrectly assumed that computing approaches – in any case not easy before the advent of the computer age – are of no real interest to the history of science.[7]

Today, through modern planetarium software, it is very easy, even for non-professionals, to reproduce Galileo's observations, at least qualitatively, and it is possible to find examples in literature and on the Web.[8,9] It is less straightforward to reconstruct with high angular precision the elongations of the satellites at the right observation time, in order to measure the effective precision and resolution of Galileo's observations. As far as I know, nobody has done it yet in a complete fashion.

Inspired by Meeus' work, I decided to analyse all of Galileo's 65 Jovian observations myself, by using tools now available to any amateur. I compared all of Galileo's angular measurements with the reconstructed elongations, obtained using *TheSky 6* Professional Edition.[10] The ad-

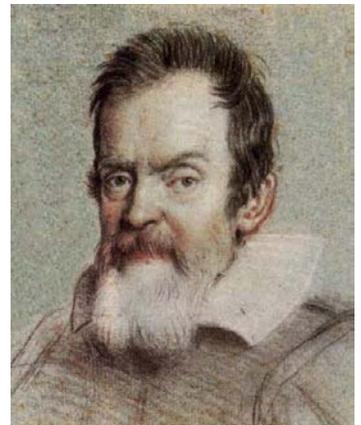

**Portrait of Galileo by Ottavio Leoni, 1624.** *Wikipedia Commons.*

vantage of using a modern and versatile software program as 'time machine' compared to Meeus' remarkable calculations is obvious. It is straightforward to move continuously in time and set any location, to obtain precise and complete sets of angular data. There are, however, some critical questions that must be answered before the simulations can be used.

First, the times reported by Galileo are not precise. Drake & Kowal, who thoroughly analysed Galileo's original manuscripts,[11] estimated that Galileo measured time with an indeterminacy of about 15 minutes. This implies, mainly for the inner (and faster) satellites, an angular indeterminacy of many arcseconds. For the 'critical' observations, near to the resolution limit, the angular positions must also be calculated 15 minutes before and 15 minutes after the time reported by Galileo, in order to obtain, instead of a point, a more correct angular *interval*.

The second question is the angular precision in the reconstructed positions obtained by the software. *TheSky 6* allows angular measurements on the screen with an indicated precision of one arcsecond. Nonetheless, how reliable is it to reconstruct satellite positions from 400 years ago? For the calculation of Jupiter satellites' positions *TheSky 6* includes software routines from the book *Astronomical Algorithms* by Meeus, that sets out the theory elaborated by the astronomer J. H. Lieske, with improvements published in 1987 known as 'E2x3'.[12] Considering that in the first Meeus' paper[1] positions are given with an accuracy of 0.1 Jupiter radii

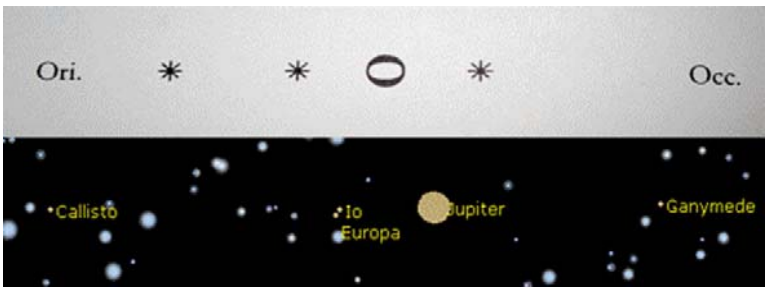

Figure 1. Galileo's drawing of the first observation of Jupiter satellites (from *Sidereus Nuncius*) and the reconstructed configuration with *The Sky6*. (Star charts generated by *TheSky6* © Software Bisque, Inc. All rights reserved. www.bisque.com.)

Table 1. Comparison between Galileo's measurements and computed data for Jupiter–satellite pairs

*(NR: Not Resolved by Galileo, I: Io, E: Europa, G: Ganymede, C: Callisto, J: Jupiter). The values in brackets are computed 15 minutes before and after the time reported by Galileo.*

| Observation no. | Jupiter–satellite (Galileo's angular estimate) | 'True' angular elongation (software reconstruction) |
|---|---|---|
| 6 | EJ (2') | 57" |
| 8 | JI NR | 32" |
| 10 | JI NR | 47" |
| 13 | IJ NR | 41" |
| 15 | EJ (1') | 50"(46"–53") |
| 18 | CJ (50") | 51"(48"–52") |
| 22 | IJ NR | 46"(42"–50") |
|    | JE NR | 49"(46"–52") |
|    | JG NR | 1'(57"–1'3") |
| 28 | JI NR | 43" |
| 30 | JI NR | 1'12"(1'9"–1'15") |
| 32 | (EI)J (20") | 52"(49"–57") |
| 41 | JI NR | 30" |
| 43 | IJ NR | 30" |
| 48 | JC NR | 52"(51"–54") |
| 49 | IJ NR | 35" |
|    | JC NR | 1'24" |
| 52 | JE NR | 46" |
| 55 | JI NR | 42" |

(about 2 arcseconds) the improved algorithms are enough to give a precision of around 1 arcsecond.

I also considered more recent results of perturbations affecting the orbit of Io that, due to the combined tidal forces of Jupiter and Europa, is the most perturbed satellite.[13] Calculations show that the overall variation in 400 years is quite small. In fact tidal forces produce a phase lag in the orbit of Io of about $4\times10^{-6}$ degree/day which implies a mismatch in the orbital position of about $0.4°$ in 400 years. This shift results in an apparent angular displacement seen from the Earth of less than an arcsecond, of the order of accuracy of the position reconstructed with the software. It is estimated that the effects of perturbation on the outer satellites are even smaller.

Once I established this important point, I began my work. I started a fascinating journey following Galileo, night by night, through 46 nights, from 1610 January 7 to March 2. In that winter, Galileo missed only nine nights due to a cloudy sky, and performed a total of 65 observations.

Galileo started his observations of Jupiter on the seventh day of January, 1610, *'the first hour of the... night'*, which is an hour after sunset. It takes Galileo four observations to realise that the 'little stars' are effectively moving around Jupiter. At the next observation, on January 12, he began to record time and elongations systematically. Only at the 6th observation did Galileo notice, for the first time, the presence of four satellites.

To reconstruct his observations, I set the software to the geographical coordinates of Padua (45°24'07" North, 11°52'06" East) and then to the day, month and year of each observation. It is important to point out, and to set the right location, that the observations from January 30 to February 12 were performed from Venice, as written in Galileo's correspondence. The software gives the time of sunset for each day and location. The convention used for the time is not important (UT, CET with daylight savings, etc.) since times are quoted relative to sunset. To reach the right time it is enough to shift the time to the hours (and/or minutes) later indicated by Galileo.

In general Galileo, on his manuscripts, counts the time either from sunset or from noon. All historical analyses have made the assumption that in the *Sidereus*, the hours (and minutes) always refer to the time after sunset. This is explicitly written by Galileo for some observations ('*ab occasu*'; see, as example, observation no.10, 1610 January 17). In the other cases this is self-evident: most of the observations start within the first hours of the night (see, as example, the first observation of January 7: '*hora sequentis noctis prima*') and a different assumption would give an improbable (or impossible) daylight observation. In addition, all the positions computed counting from sunset show a very strong resemblance with Galileo's drawings. This resemblance is completely lost assuming noon as starting time.

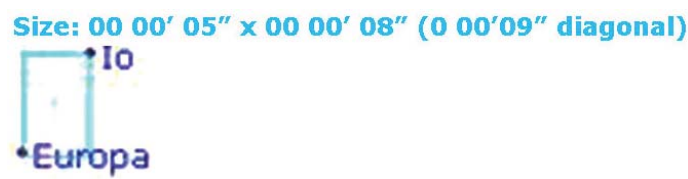

Figure 2. The angular separation between Europa and Io during Galileo's first observation. *TheSky6* © Software Bisque, Inc.

I obtained, as in previous reconstructions, a very good correlation between Galileo's drawings and the reconstructed patterns, except in one case that was not previously noted. Observation no.46 was completely discordant. This observation was made on February 12, the last day Galileo was in Venice, and was probably performed in Venice, but recorded later in Padua, reporting a wrong time and (possibly) day. However, as my preliminary intention was to perform measurements, Galileo's numbers were more relevant than his drawings.

I recorded all the angular records (more than 140) reported in the *Sidereus* and I measured all the correspondent angular separation with *TheSky 6*. In many cases, in order to obtain a more complete and relevant set of data, I also measured the angular separations *not* resolved by Galileo. Tracking on the screen the distance between two points, for example two satellites, angular separations are shown directly by the software. I used as 'start' and 'end' points the centre of satellites' disks, or the edge of the planet when Jupiter was involved. I checked the result, in some cases calculating angular separation by using celestial equatorial coordinates (RA and Dec) and the expression for angular separation θ on the sphere.* I verified that the two methods give the same results within the precision required (one arcsecond) and then used the faster and easier 'tracking' method to measure angular separations.

---

* The angular separation between two points of coordinates (RA1,Dec1) and (RA2,Dec2) is: $\cos(\theta) = \sin(Dec1)\times\sin(Dec2) + \cos(Dec1)\times\cos(Dec2)\times\cos(RA1-RA2)$

Table 2. Comparison between Galileo's measurements and computed data for satellite–satellite pairs

*(NR: Not Resolved by Galileo, I: Io, E: Europa, G: Ganymede). The values in brackets are computed 15 minutes before and after the time reported by Galileo.*

| Observation no. | Satellite–satellite (Galileo's angular estimate) | 'True' angular elongation (software reconstruction) |
|---|---|---|
| 9  | GE NR     | 11" |
| 10 | EG NR     | 16" |
| 11 | GE (20")  | 23"(19"–28") |
| 19 | IE (40")  | 20"(24"–17") |
| 22 | EG NR     | 11" |
| 24 | GE NR     | 12" |
| 29 | EG NR     | 15" |
| 40 | IE NR     | 19"(19"–19") |

An illustrative case is the first observation, shown in Figure 1, which, as noted by Meeus, includes all the four satellites. According to my measurements, as shown in Figure 2, Io and Europa are separated by 9 arcseconds. Taking into account Galileo's time indeterminacy, this value can range between 8 and 11 arcseconds. This range puts a first lower limit on the satellite to satellite resolution achieved by Galileo. But, as we will see, it is necessary to distinguish between Jupiter–satellite and satellite–satellite elongations.

Table 1 shows the most interesting cases, extracted from the complete set of Galileo's and software measurements, regarding Jupiter–satellite elongations. In all the other cases (not reported here) the angular separation Jupiter–satellite is lower than 30 arcseconds – and the satellite is not resolved – or is higher than 1 minute of arc – and the satellite is resolved.

Observations 6, 15, 18 and 32 indicate a lower limit for the Jupiter–satellites resolution of about 50 arcseconds. It is easy to see that almost all the other observations in which Galileo doesn't resolve the satellites from Jupiter, are related to smaller angular displacements, except in the case of observation 48 which is near the lower limit, and possibly, 22 (Jupiter–Ganymede) and 30 (Jupiter–Io). The last two seem to be quite discordant with the whole set of data, but it is reasonable to assume that these two exceptions are not particularly significant and could be due either to recording errors or to peculiar observing conditions. The value found for Jupiter–satellite resolution is rather smaller but consistent with Meeus' previous results.

Of course this is *not* the resolution of Galileo's telescope. The strong magnitude difference (about 7) between Jupiter and its satellites prevents us from applying the angular resolution limit usually used for visual observations based on Rayleigh's criterion: $R \approx 0.128/D$, where R is the resolution limit in arcseconds and D the objective diameter in metres.[14] The glare of Jupiter, probably increased by aberrations and glass impurities, didn't allow Galileo to resolve satellites within about two and a half equatorial Jupiter radii from the planet's edge.

From the resolution point of view, it is more significant to analyse the satellite–satellite angular separations since the magnitudes are comparable. Table 2 shows a selection of the most interesting cases. The crucial observation is no.40 of February 8, shown in Figure 3. Regarding Io and Europa, which Galileo doesn't resolve, he wrote: '...I doubt whether the closest to Jupiter was only one or two stars.' As the calculated angular separation – 19 arcseconds – doesn't change in the time error interval, it is very relevant because it indicates quite precisely the limit of resolution. This result is consistent with all the other, resolved and not resolved, satellite–satellite observations. In particular, the resolved observations nos.11 and 19 show with strong evidence and consistently with the above, a resolution of about 20 arcseconds.

This value is about twice the value hypothesised by Drake and by other historians of science and agrees with other past hypotheses based on the analysis of subsequent telescopes built by Galileo.[5,6] This is the first 'direct' measurement of resolution of Galileo's first observations, based on data from Galileo himself.

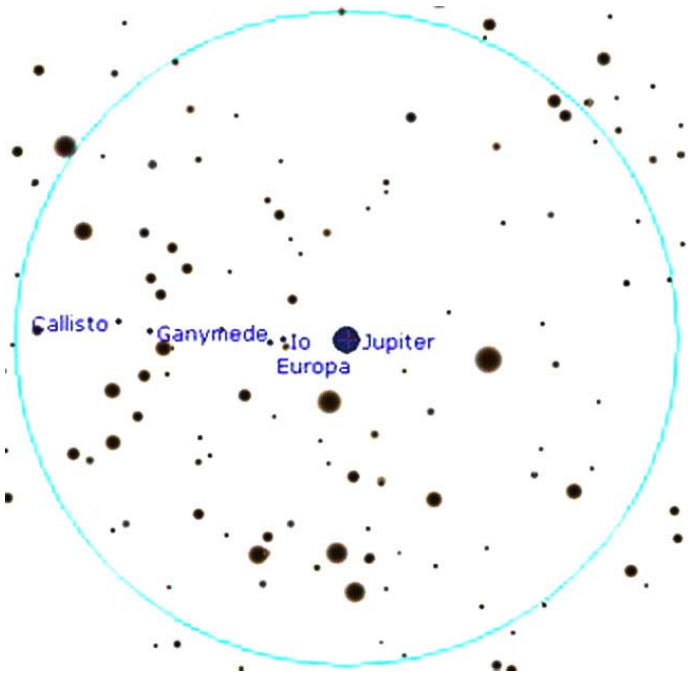

**Figure 3.** Reconstruction of observation no.40. Europa and Io (separated by 19") are reported as a single moon in Galileo's drawing, but he '...doubt whether... was only one or two stars.' The blue circle is the FOV (about 16') of a ×20 magnification Galilean telescope. (Star charts generated by *TheSky6* © Software Bisque, Inc.)

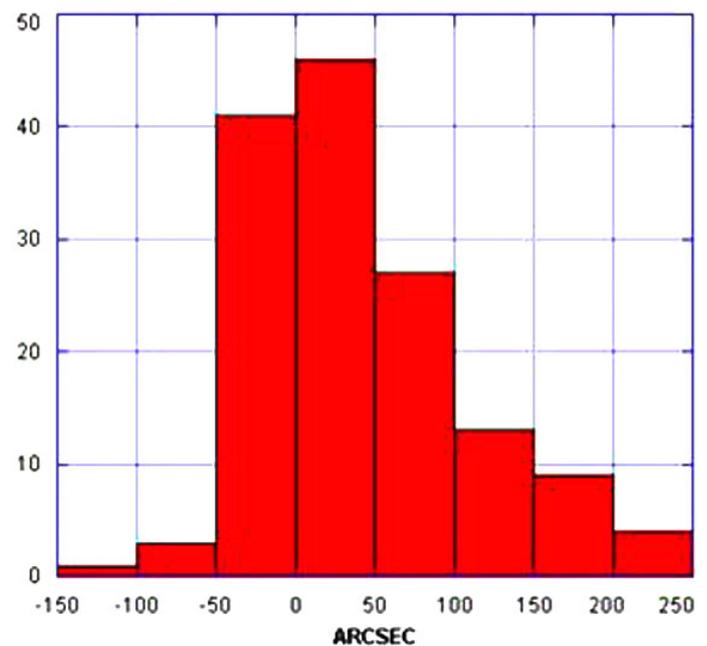

**Figure 4.** Histogram of Galileo's angular errors: errors in excess are prevalent. The errors are computed by subtracting the 'correct' angular separations, obtained from the planetarium software, from Galileo's angular measurements.

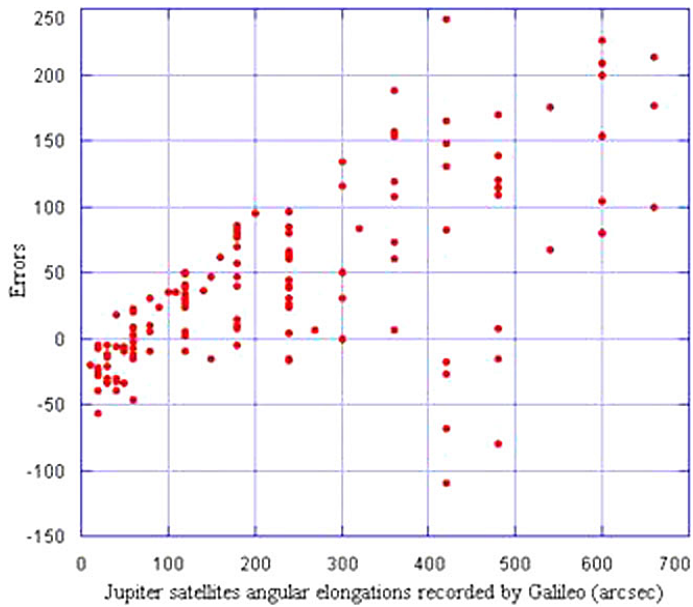

**Figure 5.** Correlation between all Galileo's angular measurements of Jupiter satellites' elongations and experimental errors. The graph shows a quite evident correlation (the correlation coefficient is about 0.7).

It is well known that Galileo probably put a diaphragm on the lens of his telescope to increase the focal ratio and reduce aberrations (mainly chromatic). The effective diameter of his lens is estimated to range between 15 and 25mm, giving a theoretical resolution limit of about 8–5 arcseconds. The actual resolution of about 20 arcseconds implies the presence of other effects degrading the resolution. The telescope's aberrations were probably added to Galileo's eye disease that he suffered when still at school and that plagued him from time to time until his later blindness.[15]

From the complete analysis of all the measurements I obtained many other interesting suggestions, either related to single observations or regarding the statistical behaviour of the data. Figure 4, in particular, shows the statistical distribution of errors in Galileo's angular measurements. The errors are obtained by comparing Galileo's measurements to the 'true' (in the meaning of 'very precise') values of the software-reconstructed angular positions. Averaging the absolute value of errors it is amazing to measure a mean error of 57 arcseconds, the same order as the one estimated by Galileo himself in the *Sidereus*: *'... unless the error of a single minute or two.'* But a quite strong *asymmetry* in the error distribution is also evident, with *positive* errors prevailing over the others.

As shown in Figure 5, the analysis of the data also shows a certain correlation between errors and measurements: the error increasing with increasing elongation. This would seem to indicate a systematic effect, that could be due to the unknown method that Galileo used to perform his angular measurements at the time of *Sidereus*' observations. Indeed, the method described by Galileo in the *Sidereus*, of putting diaphragms of varying aperture over the objective lens, cannot work, because it does not limit the field of view of the telescope, but only its resolution.[16] This is probably only Galileo's idea and not the method he effectively used. It should be remembered that *Sidereus* was written in a very short time and that Galileo did not know the laws of optics.

It is well known that *later* Galileo developed a different technique for angular measurements, reaching a remarkable precision,

and that he used the angular diameter of Jupiter as a reference scale, as he says in the *Discourse on Bodies Floating in Water*, published in 1612.[17,18,19] Giovanni Alfonso Borelli (1608–1679), Italian physiologist and physicist, describes a sort of wood ruler that Galileo is thought to have used in these first observations.[20] But this point is still unclear and is an interesting topic that requires further investigation.

Four hundred years after the publication of *Sidereus Nuncius* and 46 years after Meeus' paper, thanks to the meticulous observations of Galileo and to the software that is now readily available, I have shown that the amateur can obtain interesting and new results that can help clarify some questions that are still debated related to the first use of telescopes in astronomy.

## Acknowledgments

I would like to thank Marco Litterio for stimulating and useful discussions leading to this paper, and Prof Gheorghe Stratan who opened me to a wider vision of Galileo's work.

**Address:** INFN & Physics Dept., Università Roma Tre, Via della Vasca Navale, 84 00146 Roma, Italy. [bernieri@lnf.infn.it]